\documentclass[10pt, preprint]{sigplanconf}

\usepackage[utf8]{inputenc}
\usepackage{amsmath}
\usepackage{natbib}
\usepackage{underscore}
\usepackage{graphicx}
\usepackage{tabu}
\usepackage{pbox}
\usepackage{subcaption}
\usepackage{algorithm2e}
\usepackage{pbox}
\usepackage{titlesec}
\usepackage{float}

\newcommand{\pl}{~\text{`$+$'}~}
\newcommand{\mi}{~\text{`$-$'}~}
\newcommand{\pr}{~\text{`$*$'}~}
\newcommand{\di}{~\text{`$/$'}~}

\makeatletter
\renewcommand{\paragraph}{%
  \@startsection{paragraph}{4}%
  {\z@}{0ex \@plus .2ex}{-1em}%
  {\normalfont\normalsize\bfseries}%
}
\makeatother

\begin{document}
\toappear{}

\begingroup

% Tighten whitespace.
\titlespacing{\section}{0pt}{5pt}{-\parskip}
\titlespacing{\subsection}{0pt}{5pt}{-\parskip}
\setlength{\belowcaptionskip}{-10pt}
\setlength{\textfloatsep}{10pt plus 10pt}

\setlength{\pdfpageheight}{\paperheight}
\setlength{\pdfpagewidth}{\paperwidth}

%\titlebanner{banner above paper title}        % These are ignored unless
%\preprintfooter{short description of paper}   % 'preprint' option specified.

\preprintfooter{Parsing Expression Grammars Made Practical}

\title{Parsing Expression Grammars Made Practical}

\authorinfo{Nicolas Laurent
            \titlenote{Nicolas Laurent is a research fellow of the Belgian fund
              for scientific research (F.R.S.-FNRS).}}
            {ICTEAM, Université catholique de Louvain, Belgium}
            {nicolas.laurent@uclouvain.be}

\authorinfo{Kim Mens}
           {ICTEAM, Université catholique de Louvain, Belgium}
           {kim.mens@uclouvain.be}

\maketitle

\begin{abstract}
  \noindent Parsing Expression Grammars (PEGs) define languages by specifying a
  recursive-descent parser that recognises them. The PEG formalism exhibits
  desirable properties, such as closure under composition, built-in
  disambiguation, unification of syntactic and lexical concerns, and closely
  matching programmer intuition. Unfortunately, state of the art PEG parsers
  struggle with left-recursive grammar rules, which are not supported by the
  original definition of the formalism and can lead to infinite recursion under
  naive implementations. Likewise, support for associativity and explicit
  precedence is spotty. To remedy these issues, we introduce Autumn, a general
  purpose PEG library that supports left-recursion, left and right associativity
  and precedence rules, and does so efficiently. Furthermore, we identify infix
  and postfix operators as a major source of inefficiency in left-recursive PEG
  parsers and show how to tackle this problem. We also explore the extensibility
  of the PEG paradigm by showing how one can easily introduce new parsing
  operators and how our parser accommodates custom memoization and error
  handling strategies. We compare our parser to both state of the art and
  battle-tested PEG and CFG parsers, such as Rats!, Parboiled and ANTLR.
\end{abstract}

\category{D.3.4}{Programming Languages}{Parsing}

\keywords parsing expression grammar, parsing, \\left-recursion, associativity, precedence

%%%%%%%%%%%%%%%%%%%%%%%%%%%%%%%%%%%%%%%%%%%%%%%%%%%%%%%%%%%%%%%%%%%%%%%%%%%%%%%%
\section{Introduction}

\paragraph{Context} Parsing is well studied in computer science. There is a long
history of tools to assist programmers in this task. These include parser
generators (like the venerable Yacc) and more recently parser combinator
libraries \cite{parser-combinators}.

Most of the work on parsing has been built on top of Chomsky's context-free
grammars (CFGs). Ford's parsing expression grammars (PEGs)~\cite{peg} are an
alternative formalism exhibiting interesting properties. Whereas CFGs use
non-deterministic choice between alternative constructs, PEGs use prioritized
choice. This makes PEGs unambiguous by construction. This is only one of the
manifestations of a broader philosophical difference between CFGs and PEGs.

CFGs are generative: they describe a language, and the grammar itself can be
used to enumerate the set of sentences belonging to that language. PEGs on the
other hand are recognition-based: they describe a predicate indicating whether a
sentence belongs to a language.

The recognition-based approach is a boon for programmers who have to find
mistakes in a grammar. It also enables us to add new parsing operators, as we
will see in section \ref{features}. These benefits are due to two PEG
characteristics. (1) The parser implementing a PEG is generally close to the
grammar, making reasoning about the parser's operations easier. This
characteristic is shared with recursive-descent CFG parsers. (2) \emph{The
  single parse rule}: attempting to match a parsing expression (i.e. a sub-PEG)
at a given input position will always yield the same result (success or failure)
and consume the same amount of input. This is not the case for CFGs. For
example, with a PEG, the expression $(a*)$ will always greedily consume all the
$a$'s available, whereas a CFG could consume any number of them, depending on
the grammar symbols that follow.

\paragraph{Challenges} Yet, problems remain. First is the problem of
left-recursion, an issue which PEGs share with recursive-descent CFG parsers.
This is sometimes singled out as a reason why PEGs are frustrating to use
\cite{parsing-not-solved}. Solutions that do support left-recursion do not
always let the user choose the associativity of the parse tree for rules that
are both left- and right-recursive; either because of technical limitations
\cite{ometa-left-peg} or by conscious design \cite{direct-left-peg}. We contend
that users should be able to freely choose the associativity they desire.

Whitespace handling is another problem. Traditionally, PEG parsers do away with
the distinction between lexing and parsing. This alleviates some issues with
traditional lexing: different parts of the input can now use different lexing
schemes, and structure is possible at the lexical level (e.g. nested comments)
\cite{peg}. However, it means that whitespace handling might now pollute the
grammar as well as the generated syntax tree. Finally, while PEGs make
linear-time parsing possible with full memoization\footnote{In this context,
  memoization means caching the result of the invocation of a parsing expression
  at a given position.}, there is a fine balance to be struck between
backtracking and memoization~\cite{packrat-worth-it}. Memoization can bring
about runtime speedups at the cost of memory use. Sometimes however, the run
time overhead of memoization nullifies any gains it might bring.

Other problems plague parsing tools of all denominations. While solutions exist,
they rarely coexist in a single tool. Error reporting tends to be poor, and is
not able to exploit knowledge held by users about the structure of their
grammars. Syntax trees often consist of either a full parse tree that closely
follows the structure of the grammar, or data structures built on the fly by
user-supplied code (\emph{semantic actions}). Both approaches are flawed: a full
parse tree is too noisy as it captures syntactic elements with no semantic
meaning, while tangling grammatical constructs and semantic actions (i.e. code)
produces bloated and hard-to-read grammars. Generating trees from declarative
grammar annotations is possible, but infrequent.

\paragraph{Solution} To tackle these issues, we introduce a new parsing library
called Autumn. Autumn implements a generic PEG parser with selective
memoization. It supports left-recursion (including indirect and hidden
left-recursion) and both types of associativity. It also features a new
construct called \emph{expression cluster}. Expression clusters enable the
aforementioned features to work faster in parts of grammars dedicated to postfix
and infix expressions. Autumn also tackles, to some extent, the problems of
whitespace handling, error reporting, and syntax tree creation. By alleviating
real and significant pain points with PEG parsing, Autumn makes PEG parsing more
practical.

\paragraph{Structure} This paper starts by describing the issues that occur when
using left-recursion to define the syntax of infix and postfix binary operators
(section \ref{problem}). Next we will describe our solutions to these issues
(section \ref{implementation}). Then, we show the freedom afforded by the PEG
paradigm regarding extensibility and lay down our understanding of how this
paradigm may be extended further (section \ref{features}). Finally, we compare
Autumn to other PEG and CFG parsers (section \ref{evaluation}) and review
related work (section \ref{related}) before concluding. Because of space
restrictions, we do not review the basics of the PEG formalism, but refer to
Ford's original paper \cite{peg}.

%%%%%%%%%%%%%%%%%%%%%%%%%%%%%%%%%%%%%%%%%%%%%%%%%%%%%%%%%%%%%%%%%%%%%%%%%%%%%%%%
\section{Problems Caused by Binary Operators}
\label{problem}

\noindent This section explains why infix and postfix
binary\footnote{\textit{Binary} should be understood broadly here: $n$-ary infix
  operators (such as the ternary conditional operator) can be modelled in terms
  of binary operators.} operators are a significant pain point in terms of
expressivity, performance, and syntax tree construction. Even more so with PEGs,
due to their backtracking nature and poor handling of left-recursion. These
issues motivate many of the features supported by our parser library. Our
running example is a minimalistic arithmetic language with addition,
subtraction, multiplication and division operating on single-digit numbers.
Table \ref{tab:grammars} shows four PEGs that recognise this language, albeit
with different associativity. They all respect the usual arithmetic operator
precedence. Grammars (a), (b) and (c) are classical PEGs, whose specificities we
now examine. Grammar (d) exhibits our own \emph{expression clusters} and
represents our solution to the problems presented in this section.

\paragraph{No support for left-recursion.} The recursive-descent nature of PEGs
means that most PEG parsers cannot support all forms of left-recursion,
including indirect and hidden left recursion.\footnote{To the best of our
  knowledge, Autumn is the only available parser to support all forms of
  left-recursion with associativity selection.} Left-recursion is direct when a
rule designates a sequence of which the first element is a recursive reference,
or when a rule designates a choice which has such a sequence as alternate. The
reason this type of recursion is singled out is that it is easy to transform
into a non-left-recursive form. Left-recursion is hidden when it might or might
not occur depending on another parsing expression. For instance, $X = Y? $~$ X$
can result in hidden left-recursion because $Y?$ might succeed consuming no
input.

PEG parsers that do not support left-recursion can only handle grammars (a) and
(b). These parsers are unable to produce a left-associative parse of the input.
Some tools can handle direct left-recursive rules by rewriting them to the
idiomatic (b) form and re-ordering their parse tree to simulate a
left-associative parse \cite{rats, antlr}.

We argue it is necessary to support indirect and hidden left-recursion, so that
the grammar author is able to organise his grammar as he sees fit. Autumn
supports left-recursion natively, as will be described in section
\ref{implementation}. Using expression clusters, the associativity for operators
that are both left- and right-recursive can be selected explicitly by using the
@left_recur annotation (as shown in grammar (d)). Right-associativity is the
default, so no annotations are required in that case.

\paragraph{Performance issues in non-memoizing parsers.} Grammar (a) is parsed
inefficiently by non-memoizing parsers. Consider a grammar for arithmetic with L
levels of precedence and P operators. In our example, $L=P=2$. This grammar will
parse a single number in $O((P+1)^L)$ expression invocations (i.e. attempts to
match an expression). For the Java language this adds up to thousands of
invocations to parse a single number. The complexity is somewhat amortized for
longer arithmetic expressions, but the cost remains prohibitively high.
Memoizing all rules in the grammar makes the complexity $O(PL)$, but this
coarse-grained solution might slow things down because of the inherent
memoization overhead: cache lookups can be expensive \cite{packrat-worth-it}. In
contrast, parsing a number in grammar (b) is always $O(PL)$. Nevertheless,
grammar (a) still produces a meaningful parse if the operators are
right-associative. Not so for grammar (b), which flattens the parse into a list
of operands.

PEG parsers supporting left-recursion can use grammar (c), the layered,
left-associative variant of grammar (a). Our own implementation of
left-recursion requires breaking left-recursive cycles by marking at least one
expression in the cycle as left-recursive. This can optionally be automated. If
the rules are marked as left-recursive, using grammar (c) we will parse a
single-digit number in $O(PL)$. If, however, the cycle breaker elects to mark
the sequence expressions corresponding to each operator (e.g. $(E \pl S)$) as
left-recursive, then the complexity is \mbox{$O((P+1)^L)$}.

Expression clusters (grammar (d)) do enable parsing in $O(PL)$ without user
intervention or full memoization. This is most welcome, since the algorithm we
use to handle left-recursion does preclude memoization while parsing a
left-recursive expression.

\paragraph{Implicit precedence.} Grammars (a), (b) and (c) encode precedence by
grouping the rules by precedence level: operators in $S$ have more precedence
than those in $E$. We say such grammars are \emph{layered}. We believe that
these grammars are less legible than grammar (d), where precedence is explicit.
In an expression cluster, precedence starts at 0, the @+ annotation increments
the precedence for the alternate it follows, otherwise precedence remains the
same. It is also easy to insert new operators in expression clusters: simply
insert a new alternate. There is no need to modify any other parsing
expression.\footnote{We are talking about grammar evolution here, i.e. editing a
  grammar. Grammar composition is not yet supported by the library.}

%%%%%%%%%%%%%%%%%%%%%%%%%%%%%%%%%%%%%%%%%%%%%%%%%%%%%%%%%%%%%%%%%%%%%%%%%%%%%%%%

\begin{table}[t]
\footnotesize

\newcommand{\csetup}{
    \captionsetup{justification=centering, singlelinecheck=false, format=hang}}

\begin{subtable}[t]{0.175\textwidth}
\centering
\setlength{\belowcaptionskip}{0pt}
{$\!\begin{aligned}
E &= S \pl E ~|~ S \mi E ~|~ S \\
S &= N \pr S ~|~ N \di S ~|~ N \\
N &= [0-9]
\end{aligned}$}
\csetup\caption{Layered, right-associative}
\label{sub:a}
\end{subtable}
\hfill \vrule \hfill
\begin{subtable}[t]{0.25\textwidth}
\centering
\setlength{\belowcaptionskip}{0pt}
{$\!\begin{aligned}
% N &=    &&[0-9]
E &= S ~ (\pl E)* ~|~ S ~ (\mi E)* ~|~ S \\
S &= N ~ (\pr S)* ~|~ N ~ (\di S)* ~|~ S \\
N &= [0-9]
\end{aligned}$}
\csetup\caption{Idiomatic}
\label{sub:b}
\end{subtable}
\vspace{3mm}
\hrule
\vspace{3mm}

\begin{subtable}[t]{0.175\textwidth}
\centering
\setlength{\belowcaptionskip}{0pt}
{$\!\begin{aligned}
E &= E \pl S ~|~ E \mi S ~|~ S \\
S &= S \pr N ~|~ S \di N ~|~ N \\
N &= [0-9]
\end{aligned}$}
\csetup\caption{Layered, left-associative}
\label{sub:c}
\end{subtable}
\hfill \vrule \hfill
\begin{subtable}[t]{0.25\textwidth}
\centering
\setlength{\belowcaptionskip}{0pt}
{$\!\begin{aligned}
E &= expr \\
  &\to E \pl E &\text{@+} &~~\text{@left\_recur} \\
  &\to E \mi E & \\
  &\to E \pr E &\text{@+} &~~\text{@left\_recur} \\
  &\to E \di E & \\
  &\to [0-9] &\text{@+}
\end{aligned}$}
\csetup\caption{Autumn expression cluster}
\label{sub:d}
\end{subtable}

\caption{4 PEGs describing a minimal arithmetic language. E stands
  for Expression, S for Summand and N for Number.}
\label{tab:grammars}
\end{table}

%%%%%%%%%%%%%%%%%%%%%%%%%%%%%%%%%%%%%%%%%%%%%%%%%%%%%%%%%%%%%%%%%%%%%%%%%%%%%%%%
\section{Implementation}
\label{implementation}

\noindent This section gives an overview of the implementation of Autumn, and
briefly explains how precedence and left-recursion handling are implemented.

\subsection{Overview}

\noindent Autumn is an open source parsing library written in Java, available
online at \texttt{http://github.com/norswap/autumn}.

The library's entry points take a PEG and some text to parse as input. A PEG can
be thought of as a graph of parsing expressions. For instance a sequence is a
node that has edges towards all parsing expressions in the sequence. The PEG can
be automatically generated from a grammar file, or built programmatically, in
the fashion of parser combinators.

Similarly, parsing can be seen as traversing the parsing expression graph. The
order and number of times the children of each node are visited is defined by
the node's parsing expression type. For instance, a sequence will traverse all
its children in order, until one fails; a choice will traverse all its children
in order, until one succeeds. This behaviour is defined by how the class
associated to the parsing expression type implements the \texttt{parse} method
of the \texttt{ParsingExpression} interface. As such, each type of parsing
expression has its own mini parsing algorithm.

\subsection{Precedence}

\noindent Implementing precedence is relatively straightforward. First, we store
the current precedence in a global parsing state, initialized to 0 so that all
nodes can be traversed. Next, we introduce a new type of parsing expression that
records the precedence of another expression. A parsing expression of this type
has the expression to which the precedence must apply as its only child. Its
role is to check if the precedence of the parsing expression is not lower than
the current precedence, failing if it is the case, and, otherwise, to increase
the current precedence to that of the expression.

Using explicit precedence in PEGs has a notable pitfall. It precludes
memoization over \textit{(expression, position)} pairs, because the results
become contingent on the precedence level at the time of invocation. As a
workaround, we can disable memoization for parts of the grammar (the default),
or we can memoize over \textit{(expression, position, precedence)} triplets
using a custom memoization strategy.

\subsection{Left-Recursion and Associativity}
\label{leftrec}

\noindent To implement left-recursion, we build upon Seaton's work on the
Katahdin language \cite{katahdin}. He proposes a scheme to handle left-recursion
that can accommodate both left- and right-associativity. In Katahdin,
left-recursion is strongly tied to precedence, much like in our own expression
clusters. This is not a necessity however, and we offer stand-alone forms of
left-recursion and precedence in Autumn too.

Here also, the solution is to introduce a new type of parsing expression. This
new parsing expression has a single child expression, indicating that this child
expression should be treated as left-recursive. All recursive references must be
made to the new left-recursive expression.

Algorithm \ref{alg:left-rec} presents a simplified version of the parse method
for left-recursive parsing expressions. The algorithm maintains two global data
structures. First, a map from \textit{(position, expression)} pairs to parse
results. Second, a set of blocked parsing expressions, used to avoid
right-recursion in left-associative parses. A parse result represents any data
generated by invoking a parsing expression at an input position, including the
syntax tree constructed and the amount of input consumed. We call the parse
results held in our data structure \emph{seeds} \cite{ometa-left-peg} because
they represent temporary results that can ``grow'' in a bottom-up fashion.
Note that our global data structures are ``global'' (in practice, scoped to the
ongoing parse) so that they persist between (recursive) invocations of the
algorithm. Other implementations of the \texttt{parse} method need not be
concerned with them.

Let us first ignore left-associative expressions. When invoking a
parsing expression at a given position, the algorithm starts by looking for a
seed matching the pair, returning it if present. If not, it immediately adds a
special seed that signals failure. We then parse the operand, update the seed,
and repeat until the seed stops growing. The idea is simple: on the first go,
all left-recursion is blocked by the failure seed, and the result is our base
case. Each subsequent parse allows one additional left-recursion, until we have
matched all the input that could be. For rules that are both left- and
right-recursive, the first left-recursion will cause the right-recursion to kick
in. Because of PEG's greedy nature, the right-recursion consumes the rest of the
input that can be matched, leaving nothing for further left-recursions. The
result is a right-associative parse.

Things are only slightly different in the left-associative case. Now the
expression is blocked, so it cannot recurse, except in left position. Our loop
still grows the seed, ensuring a left-associative parse.

The algorithm has a few pitfalls. First, it requires memoization to be disabled
while the left-recursive expression is being parsed. Otherwise, we might memoize
a temporary result. Second, for left-associative expressions, it blocks all
non-left recursion while we only need to block right-recursion. To enable
non-right recursion, our implementation includes an escape hatch operator that
inhibits the blocked set while its operand is being parsed. This operator has to
be inserted manually.

%%%%%%%%%%%%%%%%%%%%%%%%%%%%%%%%%%%%%%%%%%%%%%%%%%%%%%%%%%%%
\newcommand{\br}[2]{~#1\lbrack#2\rbrack~}
\newcommand{\n}{\textnormal}

\LinesNumbered
\begin{algorithm}[t]
\footnotesize
\SetAlgoVlined
\SetKwProg{Parse}{parse}{:}{}
\SetKwBlock{Loop}{repeat}{}
\SetKw{Continue}{continue}
\SetKw{At}{at}
\SetKw{Remove}{remove}

seeds = \{\} \\
blocked = \lbrack\rbrack \\

\Parse{\n{expr:} left-recursive expression \At \n{position}}
{
    \If{\n{\br{\br{seeds}{position}}{expr}} exists}
    {
        \Return seeds[position][expr]
    }
    \ElseIf{\n{blocked} contains \n{expr}}
    {
        \Return failure
    }

    current = failure\\
    seeds[position][expr] = failure \\

    \If{\n{expr} is left-associative}
    {
        blocked.add(expr)
    }

    \Loop{
        result = parse(expr.operand) \\

        \If{\n{result} consumed more input than \n{current}}
        {
            current = result\\
            seeds[position][expr] = result\\
        } 
        \Else{
            \Remove seeds[position][expr] 
            
            \If{\n{expr} is left-associative}
            {
                blocked.remove(expr)
            }

            \Return current
        }
    
    }
}

\caption{Left-recursion and associativity handling.}
\label{alg:left-rec}
\end{algorithm}
%%%%%%%%%%%%%%%%%%%%%%%%%%%%%%%%%%%%%%%%%%%%%%%%%%%%%%%%%%%%

\subsection{Expression Clusters}
\label{clusters}

\noindent Expression clusters integrate left-recursion handling with precedence.
As outlined in section \ref{problem}, this results in a readable,
easy-to-maintain and performant construct.

An expression cluster is a choice where each alternate must be annotated with a
precedence (recall the @+ annotation from earlier), and can optionally be
annotated with an associativity. Alternates can additionally be marked as
left-associative, right-associativity being the default. All alternates at the
same precedence level must share the same associativity, hence it needs to be
mentioned only for the first alternate.

Like left-recursive and precedence expressions, expression clusters are a new
kind of parsing expression. Algorithm \ref{alg:cluster} describes the
\texttt{parse} method of expression clusters. The code presents a few key
differences with respect to the regular left-recursion parsing algorithm. We now
maintain a map from cluster expressions to their \emph{current precedence}. We
iterate over all the precedence groups in our cluster, in decreasing order of
precedence. For each group, we verify that the group's precedence is not lower
than the current precedence. If not, the current precedence is updated to that
of the group. We then iterate over the operators in the group, trying to grow
our seed. After growing the seed, we retry all operators in the group \emph{from
  the beginning}. Note that we can do away with the blocked set:
left-associativity is handled via the precedence check. For left-associative
groups, we increment the precedence by one, forbidding recursive entry in the
group. Upon finishing the invocation, we remove the current precedence mapping
only if the invocation was not recursive: if it was, another invocation is still
making use of the precedence.

%%%%%%%%%%%%%%%%%%%%%%%%%%%%%%%%%%%%%%%%%%%%%%%%%%%%%%%%%%%%

\LinesNumbered
\begin{algorithm}[t]
\footnotesize
\SetAlgoVlined
\SetKwProg{Parse}{parse}{:}{}
\SetKwBlock{Loop}{repeat}{}
\SetKw{Continue}{continue}
\SetKw{At}{at}
\SetKw{Break}{break}
\SetKw{Goto}{goto}
\SetKw{In}{in}
\SetKw{Remove}{remove}

seeds = \{\} \\
precedences = \{\} \\

\Parse{\n{expr:} cluster expression \At \n{position}}
{
    \If{\n{\br{\br{seeds}{position}}{expr}} exists}
    {
        \Return seeds[position][expr]
    }

    current = failure\\
    seeds[position][expr] = failure \\
    min\_precedence = precedences[expr] if defined, else 0 \\

    loop:
    \For{\n{group} \In \n{expr.groups}}
    {
        \If{\n{group.precedence $<$ min\_precedence}}{
            \Break
        }

        precedences[expr] = group.precedence + \\
        \Indp group.left\_associative ? 1 : 0 \\

        \Indm
        \For{\n{op} \In \n{group.ops}}
        {
            result = parse(op) \\

            \If{\n{result} consumed more input than \n{current}}
            {
                current = result\\
                seeds[position][expr] = result \\
                \Goto loop
            } 
        }
    }

    \Remove seeds[position][expr]\\

    \If{there is no other ongoing invocation of \n{expr}}{
        \Remove precedences[expr]
    }

    \Return current
}
\caption{Parsing with expression clusters.}
\label{alg:cluster}
\end{algorithm}

%%%%%%%%%%%%%%%%%%%%%%%%%%%%%%%%%%%%%%%%%%%%%%%%%%%%%%%%%%%%%%%%%%%%%%%%%%%%%%%%
\section{Customizing Parser Behaviour}
\label{features}

\subsection{Adding New Parsing Expression Types}

\noindent The core idea of Autumn is to represent a PEG as a graph of parsing
expressions implementing a uniform interface. By implementing the
\texttt{ParsingExpression} interface, users can create new types of parsing
expressions. Many of the features we will introduce in this section make use of
this capability.

\paragraph{Restrictions} The only restriction on custom parsing expressions is
\emph{the single parse rule}: invoking an expression at a given position should
always yield the same changes to the parse state. Custom expressions should
follow this rule, and ensure that they do not cause other expressions to violate
it. This limits the use of global state to influence the behaviour of
sub-expressions. Respecting the rule makes memoization possible and eases
reasoning about the grammar.

The rule is not very restrictive, but it does preclude the user from changing
the way other expressions parse. This is exactly what our left-recursion and
cluster operators do, by blocking recursion. We get away with this by blocking
memoization when using left-recursion or precedence. There is a workaround: use
a transformation pass to make modified copies of sub-expressions. Experimenting
with it was not one of our priorities, as experience shows that super-linear
parse times are rare. In practice, the fact that binary operators are
exponential in the number of operators (while still linear in the input size) is
a much bigger concern, which is adequately addressed by expression clusters.

\paragraph{Extending The Parse State} To be practical, custom parsing
expressions may need to define new parsing states, or to annotate other parsing
expressions. We enable this by endowing parsing expressions, parsers and parse
states with an \emph{extension object}: essentially a fast map that can hold
arbitrary data. There are also a few hooks to the library's internals. Our
design objective was to allow most native operators to be re-implemented as
custom expressions. Since many of our features are implemented as parsing
expressions, the result is quite flexible.

\subsection{Grammar Instrumentation}

\noindent Our library includes facilities to transform the expression graph
before starting the parse. Transformations are specified by implementing a
simple visitor pattern interface. This can be used in conjunction with new
parsing expression types to instrument grammars. In particular, we successfully
used custom parsing expression types to trace the execution of the parser and
print out debugging information.

We are currently developing a grammar debugger for Autumn and the same principle
is used to support breakpoints: parsing expressions of interest are wrapped in a
special parsing expression that checks whether the parse should proceed or pause
while the user inspects the parse state.

Transforming expression graphs is integral to how Autumn works: we use such
transformations to resolve recursive reference and break left-recursive cycles
in grammars built from grammar files.

\subsection{Customizable Error Handling \& Memoization}

\noindent Whenever an expression fails, Autumn reports this fact to the
configured error handler for the parse. The default error reporting strategy is
to track and report the farthest error position, along with some contextual
information.

Memoization is implemented as a custom parsing expression taking an expression
to memoize as operand. Whenever the memoization expression is encountered, the
current parse state is passed to the memoization strategy. The default strategy
is to memoize over \textit{(expression, position)} pairs. Custom strategies
allow using memoization as a bounded cache, discriminating between expressions,
or including additional parse state in the key.

\subsection{Syntax Tree Construction}

\noindent In Autumn, syntax trees do not mirror the structure of the grammar.
Instead, an expression can be \emph{captured}, meaning that a node with a
user-supplied name will be added in the syntax tree whenever the expression
succeeds. Nodes created while parsing the expression (via captures on
sub-expressions) will become children of the new node. This effectively elides
the syntax tree and even allows for some nifty tricks, such as flattening
sub-trees or unifying multiple constructs with different syntax. The text
matched by an expression can optionally be recorded. Captures are also
implemented as a custom parsing expression type.

\subsection{Whitespace Handling}

\noindent The parser can be configured with a parsing expression to be used as
whitespace. This whitespace specification is tied to \emph{token parsing
  expressions}, whose foremost effect is to skip the whitespace that follows the
text matched by their operand. A token also gives semantic meaning: it
represents an indivisible syntactic unit. The error reporting strategy can use
this information to good effect, for instance.

We mentioned earlier that we can record the text matched by an expression. If
this expression references tokens, the text may contain undesirable trailing
whitespace. To avoid this, we make Autumn keep track of the furthest
non-whitespace position before the current position.

%%%%%%%%%%%%%%%%%%%%%%%%%%%%%%%%%%%%%%%%%%%%%%%%%%%%%%%%%%%%%%%%%%%%%%%%%%%%%%%%
\section{Evaluation}
\label{evaluation}

%%%%%%%%%%%%%%%%%%%%%%%%%%%%%%%%%%%%%%%%%%%%%%%%%%%%%%%%%%%%
\begin{table}[]
\centering
\footnotesize
\begin{tabu}{lrrr}

Parser & Time (Single) & Time (Iterated) &  Memory \\

\noalign{\vskip 1mm}
\hline
\noalign{\vskip 1mm}    

Autumn              & 13.17 s & 12.66 s & 6 154 KB \\ 

\noalign{\vskip 1mm}
\hline
\noalign{\vskip 1mm}    

Mouse               & 101.43 s & 99.93 s & 45 952 KB \\

Parboiled           & 12.02 s & 11.45 s & 13 921 KB \\

Rats!               & 5.95 s & 2.41 s & 10 632 KB \\

\noalign{\vskip 1mm}
\hline
\noalign{\vskip 1mm}    

ANTLR v4 (Java 7)   & 4.63 s & 2.31 s & 44 432 KB \\

\end{tabu}
\caption{Performance comparison of Autumn to other PEG parsing tools as well as
  ANTLR. Measurements done over 34MB of Java code.}
\label{tab:comp}
\end{table}

%%%%%%%%%%%%%%%%%%%%%%%%%%%%%%%%%%%%%%%%%%%%%%%%%%%%%%%%%%%%

\noindent In Table \ref{tab:comp}, we measure the performance of parsing the
source code of the Spring framework ($\sim$ 34 MB of Java code) and producing
matching parse trees. The measurements were taken on a 2013 MacBook Pro with a
2.3GHz Intel Core i7 processor, 4GB of RAM allocated to the Java heap (Java 8,
client VM), and an SSD drive. The \emph{Time (Single)} column reports the median
of 10 task runs in separate VMs. The \emph{Time (Iterated)} column reports the
median of 10 task runs inside a single VM, after discarding 10 warm-up runs. The
reported times do not include the VM boot time, nor the time required to
assemble the parser combinators (when applicable). For all reported times, the
average is always within 0.5s of the median. All files are read directly from
disk. The \emph{Memory} column reports the peak memory footprint, defined as the
maximum heap size measured after a GC activation. The validity of the parse
trees was verified by hand over a sampling of all Java syntactical features.

The evaluated tools are \emph{Autumn}; \emph{Rats!} \cite{rats}, a state of the
art packrat PEG parser generator with many optimizations; \emph{Parboiled}, a
popular Java/Scala PEG parser combinator library; \emph{Mouse} \cite{mouse}, a
minimalistic PEG parser generator that does not allow memoization; and, for
comparison, \emph{ANTLR v4} \cite{antlr} a popular and efficient state of the
art CFG parser.

Results show that Autumn's performance is well within the order of magnitude of
the fastest parsing tools. This is encouraging, given that we did not dedicate
much effort to optimization yet. Many optimizations could be applied, including
some of those used in Rats! \cite{rats}. Each parser was evaluated with a Java
grammar supplied as part of its source distribution. For Autumn, we generated
the Java grammar by automatically converting the one that was written for Mouse.
We then extracted the expression syntax into a big expression cluster and added
capture annotations. The new expression cluster made the grammar more readable
and is responsible for a factor 3 speedup of the parse with Autumn (as compared
to Autumn without expression clusters).

%%%%%%%%%%%%%%%%%%%%%%%%%%%%%%%%%%%%%%%%%%%%%%%%%%%%%%%%%%%%%%%%%%%%%%%%%%%%%%%%
\section{Related Work}
\label{related}

\noindent Feature-wise, some works have paved the way for full left-recursion
and precedence handling. \emph{OMeta} \cite{ometa} is a tool for pattern
matching over arbitrary data types. It was the first tool to implement
left-recursion for PEGs, albeit allowing only right-associative parses.
\emph{Katahdin} \cite{katahdin} is a language whose syntax and semantics are
mutable at run-time. It pioneers some of the techniques we successfully
deployed, but is not a parsing tool per se. \emph{IronMeta} is a port of OMeta
to C\# that supports left-recursion using an algorithm developed by Medeiros et
al. \cite{peg-left}. This algorithm enables left-recursion, associativity and
precedence by compiling parsing expressions to byte code for a custom virtual
machine. However, Iron Meta doesn't support associativity handling.

%%%%%%%%%%%%%%%%%%%%%%%%%%%%%%%%%%%%%%%%%%%%%%%%%%%%%%%%%%%%%%%%%%%%%%%%%%%%%%%%
\section{Conclusion}

\noindent Left-recursion, precedence and associativity are poorly supported by
PEG parsers. Infix and postfix expressions also cause performance issues in
left-recursion-capable PEG parsers. To solve these issues, we introduce Autumn,
a parsing library that handles left-recursion, associativity and precedence in
PEGs, and makes it efficient through a construct called \emph{expression
  cluster}. Autumn's performance is on par with that of both state of the art
and widely used PEG parsers. Autumn is built with extensibility in mind, and
makes it easy to add custom parsing expressions, memoization strategies and
error handlers. It offers lightweight solutions to ease syntax tree
construction, whitespace handling and grammar instrumentation. In conclusion,
Autumn is a practical parsing tool that alleviates significant pain points felt
in current PEG parsers and constitutes a concrete step towards making PEG
parsing practical.

%%%%%%%%%%%%%%%%%%%%%%%%%%%%%%%%%%%%%%%%%%%%%%%%%%%%%%%%%%%%%%%%%%%%%%%%%%%%%%%%

% Fix mysterious page number on last page.
\thispagestyle{empty}

\acks \noindent We thank Olivier Bonaventure, Chris Seaton, the SLE reviewers
and our shepherd Markus Völter for their advice.

% \bibliographystyle{abbrvnat}
% \raggedright
% \bibliography{paper}

\begin{thebibliography}{12}
\softraggedright
\providecommand{\natexlab}[1]{#1}
\providecommand{\url}[1]{\texttt{#1}}
\expandafter\ifx\csname urlstyle\endcsname\relax
  \providecommand{\doi}[1]{doi: #1}\else
  \providecommand{\doi}{doi: \begingroup \urlstyle{rm}\Url}\fi

\bibitem[{A. Warth et al.}(2008)]{ometa-left-peg}
{A. Warth et al.}
\newblock {Packrat Parsers Can Support Left Recursion}.
\newblock In \emph{PEPM}, pages 103--110. ACM, 2008.

\bibitem[Becket and Somogyi(2008)]{packrat-worth-it}
R.~Becket and Z.~Somogyi.
\newblock {DCGs + Memoing = Packrat Parsing but Is It Worth It?}
\newblock In \emph{PADL}, LNCS 4902, pages 182--196. Springer, 2008.

\bibitem[Ford(2004)]{peg}
B.~Ford.
\newblock {Parsing Expression Grammars: A Recognition-based Syntactic
  Foundation}.
\newblock In \emph{POPL}, pages 111--122. ACM, 2004.

\bibitem[Grimm(2006)]{rats}
R.~Grimm.
\newblock {Better Extensibility Through Modular Syntax}.
\newblock In \emph{PLDI}, pages 38--51. ACM, 2006.

\bibitem[Hutton(1992)]{parser-combinators}
G.~Hutton.
\newblock {Higher-order functions for parsing}.
\newblock \emph{J. Funct. Program. 2}, pages 323--343, 1992.

\bibitem[Redziejowski(2009)]{mouse}
R.~R. Redziejowski.
\newblock {Mouse: From Parsing Expressions to a practical parser}.
\newblock In \emph{CS\&P 2}, pages {514--525}. {Warsaw University}, 2009.

\bibitem[{S. Medeiros et al.}(2014)]{peg-left}
{S. Medeiros et al.}
\newblock {Left Recursion in Parsing Expression Grammars}.
\newblock \emph{SCP 96}, pages 177--190, 2014.

\bibitem[Seaton(2007)]{katahdin}
C.~Seaton.
\newblock {A Programming Language Where the Syntax and Semantics Are Mutable at
  Runtime}.
\newblock Master's thesis, University of Bristol, 2007.

\bibitem[{T. Parr et al.}(2014)]{antlr}
{T. Parr et al.}
\newblock {Adaptive LL(*) Parsing: The Power of Dynamic Analysis}.
\newblock In \emph{OOPSLA}, pages 579--598. ACM, 2014.

\bibitem[Tratt(2010)]{direct-left-peg}
L.~Tratt.
\newblock {Direct left-recursive parsing expression grammars}.
\newblock Technical Report EIS-10-01, Middlesex University, 2010.

\bibitem[Tratt(2011)]{parsing-not-solved}
L.~Tratt.
\newblock Parsing: The solved problem that isn't, 2011.
\newblock URL
  \url{http://tratt.net/laurie/blog/entries/parsing_the_solved_problem_that_isnt}.

\bibitem[Warth and Piumarta(2007)]{ometa}
A.~Warth et al.
\newblock {OMeta: An Object-oriented Language for Pattern Matching}.
\newblock In \emph{DLS}, pages 11--19. ACM, 2007.

\end{thebibliography}

\endgroup

\end{document}